\documentclass[letter]{aa} 

\usepackage{graphicx}
\usepackage{txfonts}
\usepackage{natbib}
\usepackage{upgreek}
\usepackage{url}
\bibpunct{(}{)}{;}{a}{}{,}
\begin{document}


\title{Orbit of the Mercury-Manganese binary 41 Eridani
        \thanks{
        Based on observations made with ESO Telescopes at the La Silla Paranal Observatory
        under program IDs 088.C-0111, 189.C-0644, 090.D-0291, and 090.D-0917.}}

\author{C.A. Hummel\inst{1}
\and
M. Sch\"oller\inst{1}
\and
G. Duvert\inst{2}
\and
S. Hubrig\inst{3}
}

\institute{European Southern Observatory,
           Karl-Schwarzschild-Str.~2, 85748 Garching, Germany\\
           \email{chummel@eso.org}
\and
UJF-Grenoble 1/CNRS-INSU, Institut de Plan\'etologie et d'Astrophysique 
de Grenoble UMR 5274, Grenoble, France
\and
Leibniz-Institut f\"ur Astrophysik Potsdam, An der Sternwarte 16, 
14482 Potsdam, Germany
}



\abstract
{
Mercury-manganese (HgMn) stars are a class of slowly rotating
chemically peculiar main-sequence late B-type stars.  More than two-thirds of the HgMn stars are known to belong to spectroscopic binaries.
}
{
By determining orbital solutions for binary HgMn stars, we will be able to
obtain the masses for both components and the distance to the system.
Consequently, we can establish the position of both components in the
Hertzsprung-Russell diagram and confront the chemical peculiarities of the HgMn stars
with their age and evolutionary history.
}
{
We initiated a program to identify interferometric binaries in a sample
of HgMn stars, using the PIONIER near-infrared interferometer at the
VLTI on Cerro Paranal, Chile.  For the detected systems, we intend to
obtain full orbital solutions in conjunction with spectroscopic data.
}
{
The data obtained for the SB2 system 41\,Eridani allowed the determination
of the orbital elements with a period of just five~days and a semi-major
axis of under 2\,mas.  Including published radial velocity measurements,
we derived almost identical masses of $3.17\pm0.07\,M_\odot$ for the
primary and $3.07\pm0.07\,M_\odot$ for the secondary.  The measured
magnitude difference is less than 0.1\,mag.  The orbital parallax is
$18.05\pm0.17$\,mas, which is in good agreement with the Hipparcos trigonometric
parallax of $18.33\pm0.15$\,mas.  The stellar diameters are resolved
as well at $0.39\pm0.03$\,mas.  The spin rate is synchronized with the
orbital rate.
}
{}


   \maketitle

%
\section{Introduction}
\label{sec:intro}

The class of chemically peculiar stars can be roughly divided into three
subclasses: magnetic Ap and Bp stars, metallic-line Am stars, and
HgMn stars, which are late B-type stars showing extreme overabundances
of Hg (up to 6\,dex) and/or Mn (up to 3\,dex).  More than 150 stars with
HgMn peculiarity are currently known \citep{2009A&A...498..961R}.  Most of
these stars are rather young and found in young associations such as Sco-Cen, Orion
OB1, or Auriga OB1 \citep{2006A&A...449..327G,2010MNRAS.402.2539G}.
In contrast to magnetic Bp and Ap stars with large-scale organized
magnetic fields, HgMn stars exhibit strong overabundances of heavy
elements such as W, Re, Os, Ir, Pt, Au, Hg, Tl, Pb, or Bi.  To
explain the abundance anomalies in the atmospheres of HgMn stars,
radiatively driven selective diffusion is most often invoked. However,
since parameter-free diffusion models appear unable to reproduce
the observed abundance anomalies, other mechanisms are expected to
contribute, such as anisotropic stellar winds, membership in binary
and multiple systems, and the presence of tangled magnetic fields
\citep[e.g.,][]{1995ComAp..18..167H,2012A&A...547A..90H}.

Although strong large-scale magnetic fields have not generally
been found in  HgMn  stars,  it  has  never  been  ruled  out  that
these stars might have tangled magnetic fields with only very weak net
longitudinal components \citep{2010MNRAS.408L..61H,2012A&A...547A..90H}.
The studies of Hubrig et al.\ also suggested the  existence  of intriguing
correlations between magnetic field, abundance anomalies, and binary
properties.  In the two double-lined spectroscopic binary (SB2)
systems with synchronously rotating components, 41 Eri and AR Aur, the
stellar surfaces facing the companion star usually display low-abundance
element spots and negative magnetic field polarity.  The surface of the
opposite hemisphere, as a rule, is covered by high-abundance element
spots and the magnetic field is positive at the rotation phases of the
best-spot visibility \citep{2012A&A...547A..90H}.

Another important distinctive feature of these stars is their slow
rotation \citep[$\langle v\,\sin i\rangle \approx$ 29\,km\,s$^{-1}$;][]
{1972ApJ...175..779A}.  The number of HgMn stars decreases sharply with
increasing rotational velocity \citep{1974ApJ...194...65W}.  Evidence that
stellar rotation affects abundance anomalies in HgMn stars is provided
by the rather sharp cutoff in these anomalies at a projected rotational
velocity of 70--80\,km$\,$s$^{-1}$ \citep{1996A&AS..120..457H}.

Previous studies clearly showed that the phenomenon of late B-type
stars exhibiting HgMn abundance anomalies is intimately linked
with their multiplicity.  Some HgMn stars even belong to triple
or quadruple systems \citep{1991PASAu...9..270I,1992AJ....103.1357C}.

In agreement with previous findings, in a most representative multiplicity
study with NACO, \citet{2010A&A...522A..85S} found several so far
undetected companions around HgMn stars, pushing the number of HgMn
stars without companions in their sample well below the 10\% level.

The existence of a companion is a crucial piece of evidence that can
explain the slowdown via tidal interaction of the typical high spin rates
of early type stars after contraction to the main-sequence.  The abundance
anomalies found in the atmospheres of chemically peculiar stars are then
thought to be caused by gravitational settling or radiatively driven
diffusion in a stable environment as these stars do not have a convective
outer region and rotate slowly, avoiding meridional flows.

The high occurrence of spectroscopic binaries among the HgMn stars could,
on the other hand, just reflect the higher multiplicity rate of early type
stars \citep{2014ApJS..215...15S,2010A&A...522A..85S} compared to late
type stars as a consequence of the formation of massive stars. However,
no fast rotators are known to exhibit chemical peculiarities.

Optical interferometry in the visual and near-infrared bands is a
technique complementary to Doppler spectroscopy for the study of these
stars and allows the determination of fundamental stellar properties
including flux ratios, multiplicity rates, and surface imaging
\citep{2002AN....323..241W,2014MNRAS.443.1629S}.

Interferometric orbits were already measured for the HgMn binaries
$\alpha$\,Andromedae \citep{1992ApJ...384..624P,2004A&A...416..661C},
$\phi$\,Herculis \citep{2007ApJ...655.1046Z,2007AJ....133.2684T}, and
$\chi$\,Lupi \citep{2013A&A...551A.121L}. These resulted in mass estimates
for the HgMn primaries of $3.5\pm1.0\ M_\odot$ (B8IV), $3.05\pm0.24\
M_\odot$ (B8V), and $2.84\pm0.12\ M_\odot$ (B9.5V), respectively.  Here we
report on interferometric observations that resolved the orbit of the
HgMn star 41\,Eridani (HD 27376, HIP 20042) and resulted in the most precise 
mass estimate of a HgMn star so far.

\section{Observations and data reduction}

Observations were carried out with the PIONIER
beam combiner \citep{2011A&A...535A..67L} mounted on the VLTI
\citep{2007NewAR..51..628S}, using stations A1, G1, J3 (I1 in Jan 2012),
and K0, providing baseline lengths of up to 130 m (corresponding
to a fringe spacing of 2.5 mas).  In 2012 and 2013, fringes were
dispersed with a grism and resulted in measurements of the visibility
amplitude and closure phase in three narrowband channels
($\Delta\lambda=0.09$ nm) centered
at $1.60\ \mu$m, $1.69\ \mu$m, and $1.77\ \mu$m.  In 2014, the grism
was not used and the fringes were recorded in a single broadband channel
approximately $0.24\ \mu$m wide (details in the next section).  
The 2014 data were reduced with
the {\ttfamily pndrs} package \citep{2011A&A...535A..67L}, while the
older data, already reduced and calibrated with {\ttfamily pndrs}, 
were retrieved from the
OiDB\footnote{\url{http://oidb.jmmc.fr/index.html}} database hosted
at the Jean-Marie Mariotti Center (JMMC) in Grenoble, France. The
dates of observation and number of visibility measurements are given
in Table~\ref{tab:astro}.

\begin{table*}
\caption{PIONIER results for 41 Eridani.}
\label{tab:astro}
\centering
\begin{tabular}{lcccrrrrrrr}
\hline
\hline
\multicolumn{1}{c}{UT Date} &
\multicolumn{1}{c}{HJD $-$} &
\multicolumn{1}{c}{Number of} &
\multicolumn{1}{c}{Dispersive} &
\multicolumn{1}{c}{$\rho$} &
\multicolumn{1}{c}{$\theta$} &
\multicolumn{1}{c}{$\sigma_{\rm maj}$} &
\multicolumn{1}{c}{$\sigma_{\rm min}$} &
\multicolumn{1}{c}{$\phi$ } &
\multicolumn{1}{c}{$O-C_\rho$} &
\multicolumn{1}{c}{$O-C_\theta$} \\
 &
\multicolumn{1}{c}{$2,400,000$} &
\multicolumn{1}{c}{visibilities} &
\multicolumn{1}{c}{element} &
\multicolumn{1}{c}{(mas)} &
\multicolumn{1}{c}{(deg)} &
\multicolumn{1}{c}{(mas)} &
\multicolumn{1}{c}{(mas)} &
\multicolumn{1}{c}{(deg)} &
\multicolumn{1}{c}{(mas)} &
\multicolumn{1}{c}{(deg)} \\
\multicolumn{1}{c}{(1)} &
\multicolumn{1}{c}{(2)} &
\multicolumn{1}{c}{(3)} &
\multicolumn{1}{c}{(4)} &
\multicolumn{1}{c}{(5)} &
\multicolumn{1}{c}{(6)} &
\multicolumn{1}{c}{(7)} &
\multicolumn{1}{c}{(8)} &
\multicolumn{1}{c}{(9)} &
\multicolumn{1}{c}{(10)} &
\multicolumn{1}{c}{(11)} \\
\hline
2012 Jan 02&55928.71& 90&Grism& 1.68& 312.58& 0.04& 0.02&161.5&  0.03&   0.3\\
2012 Oct 10&56210.71& 34&Grism& 1.78& 219.78& 0.03& 0.01&129.7& $-$0.01& 0.3\\
2012 Nov 22&56253.71& 34&Grism& 1.63&   8.93& 0.03& 0.02&133.4& $-$0.00& 0.0\\
2012 Nov 23&56254.71& 34&Grism& 1.77& 287.17& 0.03& 0.02&115.8&  0.00& $-$0.6\\
2012 Nov 24&56255.71& 34&Grism& 1.81& 225.02& 0.03& 0.02&116.5& $-$0.01&$-$0.5\\
2012 Nov 25&56256.71& 34&Grism& 1.62& 145.47& 0.03& 0.02&115.0&  0.02& $-$1.5\\
2012 Nov 26&56257.71& 34&Grism& 1.92&  77.36& 0.03& 0.02&135.0&  0.02&  0.8\\
2013 Jan 24&56316.71& 34&Grism& 1.59& 156.27& 0.03& 0.02&126.2&  0.01& $-$1.0\\
2014 Aug 20&56889.71&  90&Free& 1.74&  33.18& 0.03& 0.01&112.2& $-$0.01&  0.6\\
2014 Aug 25&56894.71&  30&Free& 1.74&  33.10& 0.03& 0.02&118.9& $-$0.02&$-$0.2\\
\hline
\end{tabular}
\flushleft
Note: Astrometric positions given in columns 5 and 6 are for local 
midnight on the date of observation (HJD for 5 UT).
\end{table*}

\section{Calibration and analysis}

Calibrator stars were selected from the catalog of
\citet{2010yCat.2300....0L}.  Inspection of the transfer function (i.e.,
the visibility measured on the calibrators corrected for their finite
angular diameters) led to the exclusion of a few of the 
calibrator observations if the resulting transfer function
was too low.

\begin{figure}
\resizebox{\hsize}{!}{\includegraphics{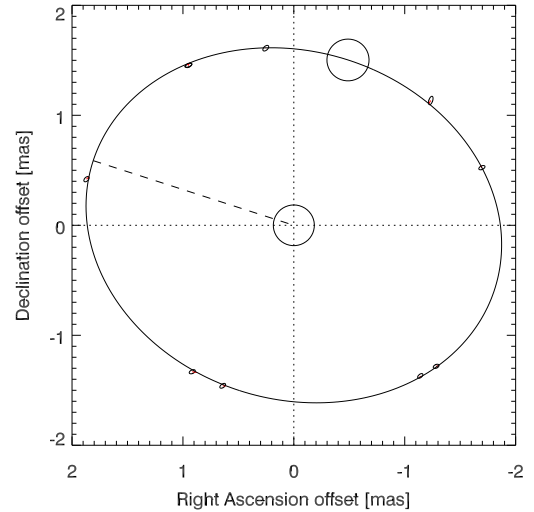}}
\caption{
Apparent orbit of 41 Eridani. The dashed line indicates the
ascending node. The small circles outline the stellar disks. 
The short (red) lines connect the measurements with the orbital model values.
}
\label{fig:orbit} 
\end{figure} 

A good knowledge of the central wavelengths for the broadband channel is
essential in order to avoid systematic errors. Model visibilities have
to be integrated over the bandpass before comparison with the observed
visibilities. The pipeline uses scans of the fringes obtained with an
internal lamp to compute the fringe amplitude power-spectral-density
distribution (PSD).  The measurement on-sky would be even better, however, the strong fringe motion blurs the PSD in this case.  For our 2014 data,
which were taken five days apart in two nights, the central wavelengths/FWHM were determined for each baseline
to be 1685/278 nm (A1-G1),
1734/146 nm (A1-J3), 1648/266 nm (G1-J3) 1679/278 nm (K0-A1), 1722/210 nm
(K0-G1), and 1685/270 nm (K0-J3).


Modeling of the (squared) visibility amplitudes and closure phases was
performed with OYSTER\footnote{\url{http://www.eso.org/~chummel/oyster}}.
Initial estimates for the location of the companion for each night were 
obtained with 
CANDID\footnote{\url{https://github.com/amerand/CANDID}}
\citep{2015A&A...579A..68G} and used to compute an initial orbit. Final
values for the orbital elements were then found by fitting all visibility
data. This fit also included as parameters the (uniform disk) sizes
of the two stars and their magnitude difference.  Limb darkening
of hot stars in the H band is not significant when compared to our
measurement uncertainties. Following \citet{2012A&A...547A..90H},
we adopted a circular orbit and achieved a reduced $\chi^2=1.5$.
Results for the fitted parameters are listed in the top
section of Table~\ref{tab:elements}. The value of the period was
refined by including the radial velocities published by \citet[
Table~I]{1915LicOB...8..168P}. The value of the periastron angle
was set to $90^\circ$ to make the epoch refer to conjunction and
required a very small adjustment of 0.0013 days to the value quoted by
\citet{2012A&A...547A..90H}.

\begin{figure}
\resizebox{\hsize}{!}{\includegraphics{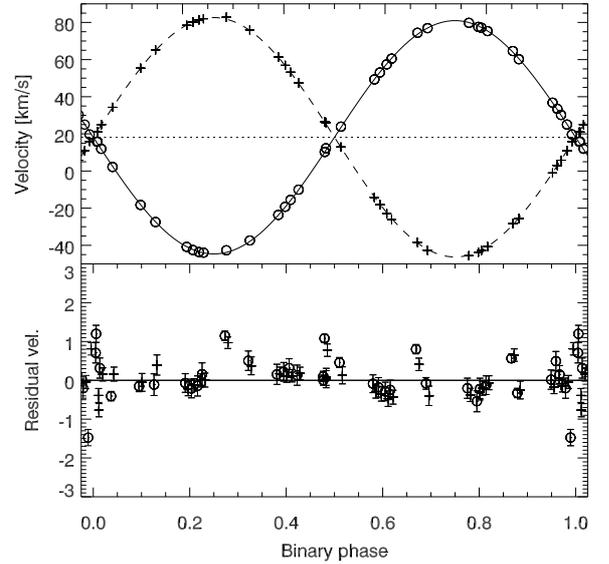}}
\caption{
Radial velocities from \citep{2012A&A...547A..90H}
for components A (circles) and B (plus symbols) 
with the solid line showing the orbital model for A and the dashed line
for B.  The lower panel shows the fit residuals.
}
\label{fig:vel} 
\end{figure}

The component masses $M$ 
(see middle section of Table~\ref{tab:elements})
could now be computed from the values of
$M\sin^3{i}$ given in \citet[Table 8]{2012A&A...547A..90H} and, using 
Kepler's third law, resulted in a parallax of $18.05\pm0.17$
mas in good agreement with the Hipparcos value of $18.33\pm0.15$
mas \citep{2007A&A...474..653V}.
The radial velocity data of \citet[Table 7]{2012A&A...547A..90H} 
are shown in Fig.~\ref{fig:vel} with our orbital model.
The increased scatter of the residuals maybe due to the presence
of the chemical spots, which can cause line profile variations (F.\
Gonzalez, priv. comm.).

To visualize the constraints provided by the interferometric observations,
the data of each night were fit with relative astrometric positions (using
the values for component diameters and relative flux from the all-night
fit). The results are plotted on the orbit in Fig.~\ref{fig:orbit}.
More details are given in Table~\ref{tab:astro} and include separation
$\rho$ and position angle $\theta$ (columns 5 and 6), the semi-major
axes of the astrometric uncertainty ellipses (in columns 7 and 8),
the position angle of the major axis (column 9), and the offsets
of the astrometric positions from the orbit in separation and angle
(columns 10 and 11). The astrometric uncertainty ellipses correspond in
size to the synthesized point spread functions (based on the achieved
interferometric aperture coverage) divided by 40 to normalize the total
astrometric reduced $\chi^2$ to unity. 
The scaling factor is a function of the uncertainties of the visibility 
and closure phase measurements, which were similar for all data sets.


\begin{table}
\caption{Orbital elements and component parameters
for 41 Eridani.
} 
\label{tab:elements}
\centering
\begin{tabular}{lr@{\,$\pm$\,}l}
\hline
\hline
Semi-major axis /mas                            & $1.902$ & $0.006$\\
Inclination $/^\circ$                   & $146.2$ & $0.1$\\
Ascending node $/^\circ$ (J2000.0)      & $72.0$ & $0.4$\\
Eccentricity                            & \multicolumn{2}{c}{$0$ (fixed)}\\
Periastron angle /$^\circ$ (primary)    & \multicolumn{2}{c}{$90$ (fixed)}\\
Periastron epoch (JD)           & $2454407.7214$ & $0.002$ \\
Period /days                    & $5.0103250$ & $0.0000008$\\
$\Delta H$                      & $0.052$ & $0.006$\\
$D_{\mathrm A,B} /$mas          & $0.39$ & $0.03$\\
\hline
$M_{\mathrm A}/M_\odot$         & $3.17$ & $0.07$\\
$M_{\mathrm B}/M_\odot$         & $3.07$ & $0.07$\\
$\pi_{\mathrm orb.} /$mas       & $18.05$ & $0.17$\\
\hline
Radius /$R_\odot$               & $2.32$ & $0.18$\\
$\log{g}$                       & $4.21$ & $0.07$\\
Luminosity (primary) /$L_\odot$ &       $100.6$ & $4.3$\\
Luminosity (secondary) /$L_\odot$&      $87.4$  & $3.3$\\
\hline 
\end{tabular}
\end{table}

\section{Discussion}

\begin{figure}
\resizebox{\hsize}{!}{\includegraphics[angle=-90]{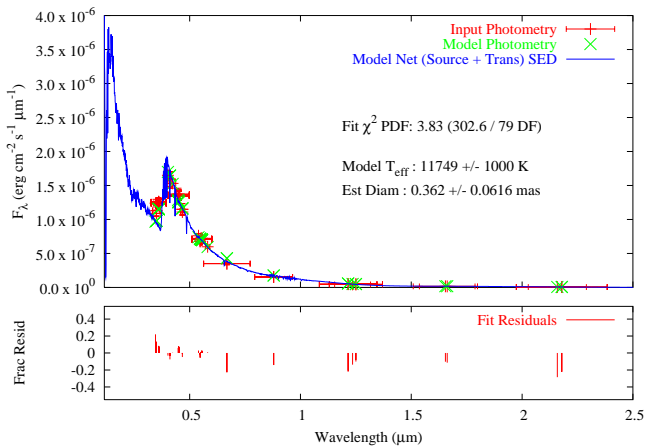}}
\caption{
Fit of apparent diameter (and effective temperature) to Simbad photometry
with {\ttfamily sedFit}. The red symbols correspond to the photometric
data and the green crosses to the model values. The blue line denotes the
template from \citet{1998PASP..110..863P} for a B8 main-sequence star.
The (reduced) $\chi^2$ of the fit is given in the figure as well.
The lower panel shows the fit residuals as fractions of the total flux.
}
\label{fig:sedfit} 
\end{figure} 

According to \citet{1915LicOB...8..168P}, 41 Eridani ``was
placed in the spectral Class B9A\footnote{The letter A was used by
\citet{1901AnHar..28..129C} to indicate the adjacent spectral class.} in
the catalogue of southern spectra given in {\it Harvard Annals}, Vol.\
28, Part 2.'' \citet{1915LicOB...8..168P} discusses the classification
based on the fact that the star still shows Helium lines, while at the
same time displaying significantly more prominent metal lines compared to
$\lambda$ Centauri, a ``typical'' star of class B9.    Initially, as the two components of 41 Eri appear nearly identical, we adopted the same
spectral type of B9V for both of them in order to obtain photometric
estimates for the apparent diameters of the stars. We used the {\ttfamily sedFit}
\citep{2007NewAR..51..617B} tool to retrieve photometric data from the
Simbad database, dividing the fluxes in half, and to fit stellar template
spectra from \citet{1998PASP..110..863P} to derive the bolometric flux
(Fig.~\ref{fig:sedfit}). The fit was significantly better with a template
of a B8V star. Using equation 6 of \citet{2007NewAR..51..617B},
$\theta^2=4F_{\rm Bol}/\sigma T^4_{\rm eff}$, with the Boltzmann
constant $\sigma$, an angular diameter $\theta=0.36\pm0.06$ mas is fitted
with a bolometric flux $F_{\rm Bol}=8.31\cdot 10^{-7}$ erg cm$^{-2}$ $s^{-1}$,
based on an effective temperature of the template of $11750\pm1000$ K.
This value is consistent with the diameter fit to the interferometric
data (given in Table~\ref{tab:elements}). 

The effective temperatures derived from spectroscopy by
\citet{2003A&A...402..299D}, $T_{\rm eff, A}=12750$ K and $T_{\rm eff,
B}=12250$ K, uncertainty of 200 K, are higher still and would correspond
to spectral types between B8 and B7 \citep{1982lbg6.conf.....A}.
Owing to the smaller uncertainty of the spectroscopic effective
temperatures, we adopt these in the following. Using again
equation 6 of \citet{2007NewAR..51..617B}, $\theta=8.17\,{\rm mas}
\cdot 10^{-0.2(V+BC)} (T_{\rm eff}/5800\,{\rm K})^{-2}$, nearly
identical angular diameters of $0.35\pm0.01$ mas are obtained
for each component with the bolometric corrections $BC_{\rm A}=-0.84$
and $BC_{\rm B}=-0.73$ \citep{1996ApJ...469..355F,2010AJ....140.1158T}.  
Here we adopted
the same magnitude difference in the $V$ band as measured in the $H$
band by PIONIER to compute individual component magnitudes from
the combined $V=3.56$ (Simbad).  The corresponding luminosities
are given in Table~\ref{tab:elements} together with the $\log{g}$
values, which are computed from the masses and absolute stellar radii
of $2.32\pm0.18\,R_\odot$. \citet{2003A&A...402..299D} derived $\log{g}=4.18\pm0.10$ for the
primary and $\log{g}=4.10\pm0.10$ for the secondary from uvby
photometry; the latter value is smaller
owing to the lower $T_{\rm eff}$.

\begin{figure}
\resizebox{\hsize}{!}{\includegraphics{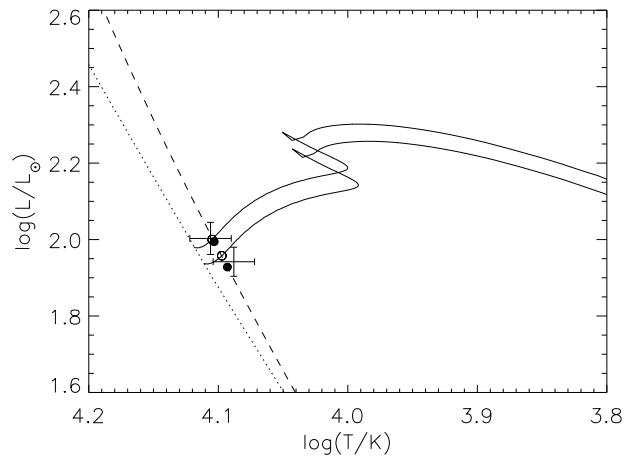}}
\caption{
Stellar evolutionary tracks (solid lines) for solar composition
stars of $3.15\ M_\odot$ and $3.07\ M_\odot$, based on models by
\citet{2012A&A...537A.146E}. The two large symbols represent the
observations with error bars in luminosity and effective temperature,
while open circles indicate the models closest to the observations.
The dashed line is an isochrone for the age of 50 My with solid circles
indicating models closest to the observed stars.  The dotted line is
the zero-age main sequence.
}
\label{fig:tracks} 
\end{figure}

We show in Fig.~\ref{fig:tracks} the stellar
evolutionary tracks of two stars with solar metallicity interpolated
for masses of the primary and secondary of 41 Eridani using models
of \citet{2012A&A...537A.146E}. The average age of the model stars
matching all observables best is $50\pm2$\,Myrs, and their average
radius is $2.1\pm0.1\,R_\odot$, which is consistent with the values quoted in
Table~\ref{tab:elements}.  The stars are therefore slightly evolved
and the average model surface gravity is $\log{g}=4.3$.  

The rotational rate of the stars was measured to be the same
($v\sin{i}=12$ km/s) by \citet{2003A&A...402..299D} and translates into
an equatorial velocity of 22 km/s (assuming spin-orbit alignment). If we
divide the stellar circumference derived from the angular diameter
and the distance ($10.1\pm0.8\cdot10^6$ km)
by the orbital period, we obtain an equatorial velocity of $23.4\pm1.8$
km/s, which is consistent with the value derived from the rotational rate
and therefore indicates synchronous corotation of the stars.


\section{Summary}

Observations of the spectroscopic binary 41 Eridani were performed
with the PIONIER H-band beam combiner at the VLTI and allowed the
component separation to be resolved at each of the 10 epochs thanks to
the high-precision data delivered by this instrument. The elements of
the five-day apparent orbit were derived and, when combined with those of
the spectroscopic orbit \citep{2012A&A...547A..90H}, delivered the stellar
masses with an accuracy of 2\% and the orbital parallax with an accuracy
of 1\%. The diameters of the nearly identical components, while barely
resolved, were consistent with estimates based on fitting the spectral
energy distribution. The derived effective temperature was higher than
expected for a main-sequence star with spectral classification B9,
confirming results by \citet{2003A&A...402..299D}, and might be related
to an erroneous classification in the {\it Harvard Annals} due to the
stronger metal lines of this chemically peculiar star.

\begin{acknowledgements}      
The authors want to warmly thank all the people involved in the
VLTI project and Jean-Baptiste Le Bouquin and the PI Team for their
contributions to PIONIER and the science-grade pipeline.
This article made use of the Smithsonian/NASA Astrophysics Data System
(ADS), of the Centre de Données astronomiques de Strasbourg (CDS), and
of the Jean-Marie Mariotti Center (JMMC).  We thank Gerard van Belle
for providing an installation of {\tt sedFit}.  We thank the anonymous
referee for comments that helped improve our paper.
\end{acknowledgements}
 
%
\bibliographystyle{aa} 
\bibliography{references} 
%

\end{document}